# Towards Scientific Literacy in Inclusive Science Education

A Self-Study Manual to Support Pre- and In-Service Teachers

Stefanie Lenzer[1,#,*], Laura Pannullo[2,#,*], Andreas Nehring[3] & Lisa Stinken-Rösner[2]

[1] IPN – Leibniz Institute for Science and Mathematics Education Kiel
[2] University Bielefeld
[3] Leibniz University Hannover
[#] These authors contributed equally to this work.
[*] Correspondence: lenzer@leibniz-ipn.de, laura.pannullo@physik.uni-bielefeld.de

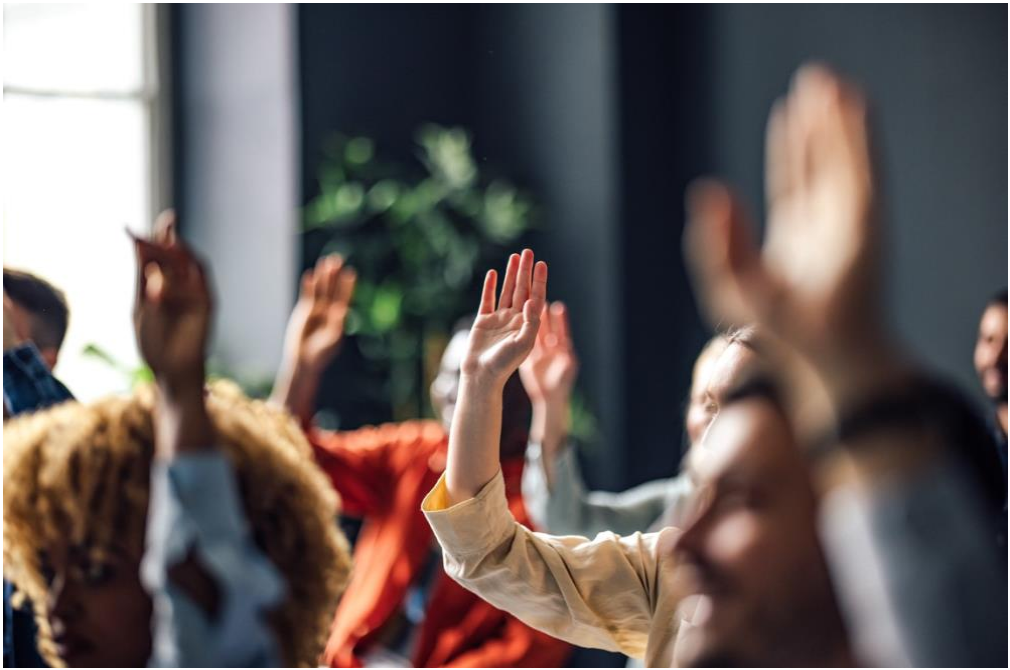

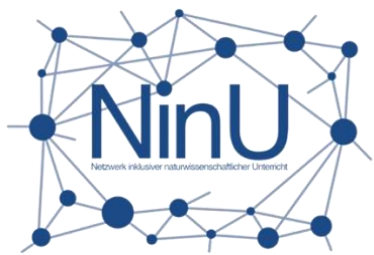

# Abstract

Scientific literacy, a central goal of modern science education, should be accessible to all students regardless of their backgrounds. Despite international education reforms focused on diversity, equity and inclusion many teachers struggle to create inclusive science lessons. One reason may be, that creating inclusive science lessons is challenging: it requires teachers to balance the demands of science education with the diverse needs of their students. Therefore, teachers need support in planning inclusive science lessons. To address this issue, members of a German network for inclusive science education (*Netzwerk inklusiver naturwissenschaftlicher Unterricht*, *NinU*) have developed the *NinU framework*. This framework integrates perspectives from both science education and inclusive pedagogy to support teachers in planning and reflecting inclusive science lessons.

Originally created as a workshop resource for the NARST 2023 conference, this self-study manual provides 1) an overview of the theory behind the *NinU framework*, 2) examples and 3) exercises for acknowledging diversity, recognizing barriers and enabling participation using the framework.

TABLE OF CONTENTS

# I. Introduction

*Scientific literacy* as a central goal of modern science education must be accessible for all students, regardless of their backgrounds. However, despite the great importance and current reforms towards equity in many countries, implementing inclusive science teaching in classroom practice remains challenging and many teachers struggle to make their lessons inclusive and accessible to all students. To address inclusion and equity in their classrooms, teachers need to know both the demands of science education and the demands of a diverse student body (Stinken-Rösner et al., 2020). They also need to know how to fulfil these two requirements at the same time. So far, teachers report missing training and lacking competencies in planning and reflecting inclusive science lesson.

Therefore, additional resources are needed to support pre-service and in-service science teachers in planning and reflecting on lessons that acknowledge students' diversity, recognize possible barriers, and enable participation for *all*.

Members of the *Network for Inclusive Science Education* (*NinU*), a collaboration of German science education researchers and practitioners, have jointly developed a framework-based approach to fill this gap. With the so called *NinU-framework* (Stinken-Rösner et al., 2020), they have succeeded in systematically combining the perspectives of science education and inclusive pedagogy to provide a resource to support pre-service and in-service science teachers in planning and reflecting on inclusive science teaching – as demonstrated in several publications for classroom practice (e.g. Ferreira González et al., 2021; Stinken-Rösner & Hofer, 2022; Watts & Weirauch, 2022). The framework is intended as a supplementary tool for lesson planning.

The examples presented in this self-study manual are based on the article by Ferreira González et al. (2021). The self-study manual was originally intended as a manual to accompany a workshop at the NARST conference in Chicago in 2023. Following the workshop, the manual was revised and now serves as a self-study manual.

## II. The NinU-framework

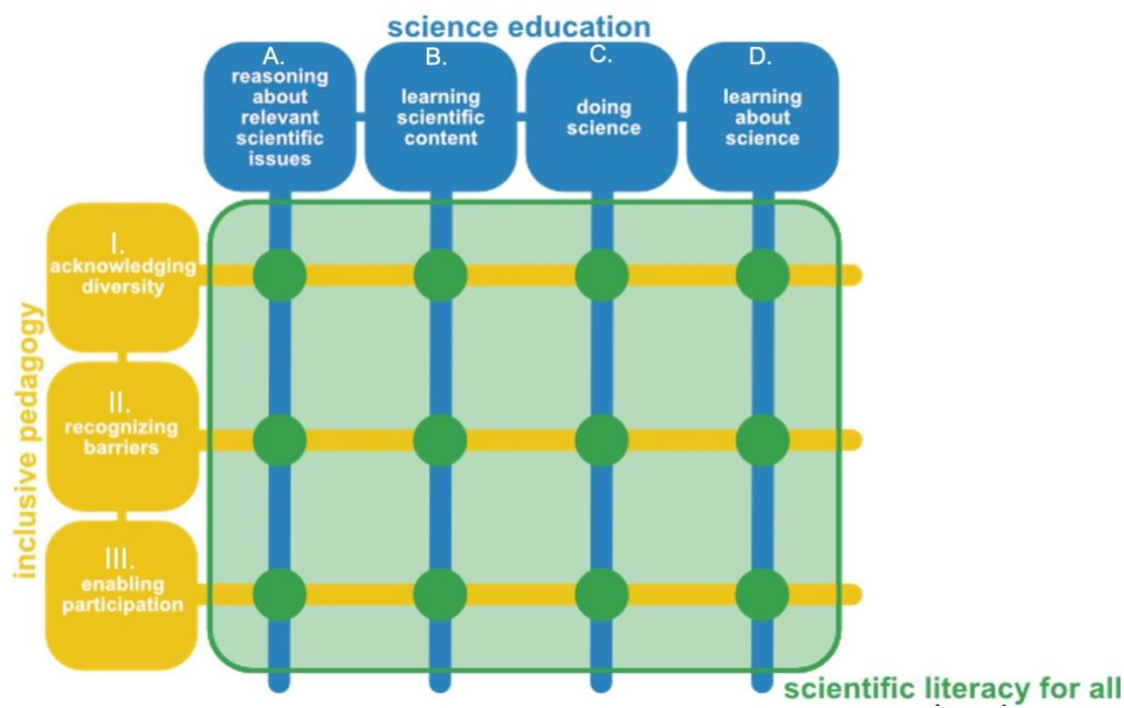

*Figure 1: The NinU-framework (adapted from Stinken-Rösner et al., 2020).*

The NinU-framework seeks to provide a new perspective on inclusive science education by systematically combining four major learning goals in science education (reasoning about relevant scientific issues, learning scientific content, doing science, and learning about science, cf. Hodson, 2014) with three dimensions of inclusive pedagogy (acknowledging diversity, recognizing barriers, and enabling participation, Fig. 1; Stinken-Rösner et al., 2020). To achieve the primary objective of promoting scientific literacy for all, the learning goals in science education (blue, Fig. 1), the dimensions of inclusive pedagogy (yellow, Fig. 1) need to be understood as interwoven. Each subcategory of science education is linked to each subcategory of inclusive pedagogy, and vice versa, as noted by Stinken-Rösner et al. (2020). To support the planning and reflection of inclusive science education lessons, each of these categories (green hubs, cf. Fig. 1) contains several guiding questions. Therefore, the NinU-framework can be used as an additional resource for teachers to plan, deliver, and reflect on science lessons that contribute to the goal of scientific literacy for all. For further information and insights into the NinU-framework, refer to Stinken-Rösner et al., 2020.

## III. How to use the NinU-framework

In applying the framework, the three dimensions of inclusive pedagogy (cf. Fig. 1: I, II, II) build structurally upon each other. Thus, addressing the dimension of enabling participation only makes sense if existing barriers have been recognized beforehand. The recognition of barriers is in turn based on the acknowledgement of diversity. Therefore, it is important to follow the three steps I-III in the given order when using the NinU-framework. For each of the four learning goals of science education (cf. Fig. 1: A, B, C, D) users are provided with guiding questions related to the dimensions of inclusive pedagogy. For all columns, it is recommended to start with the first guiding question in each column and then work through the questions to the bottom. In addition, it is recommended to proceed with the columns as described below and illustrated in Figure 2.

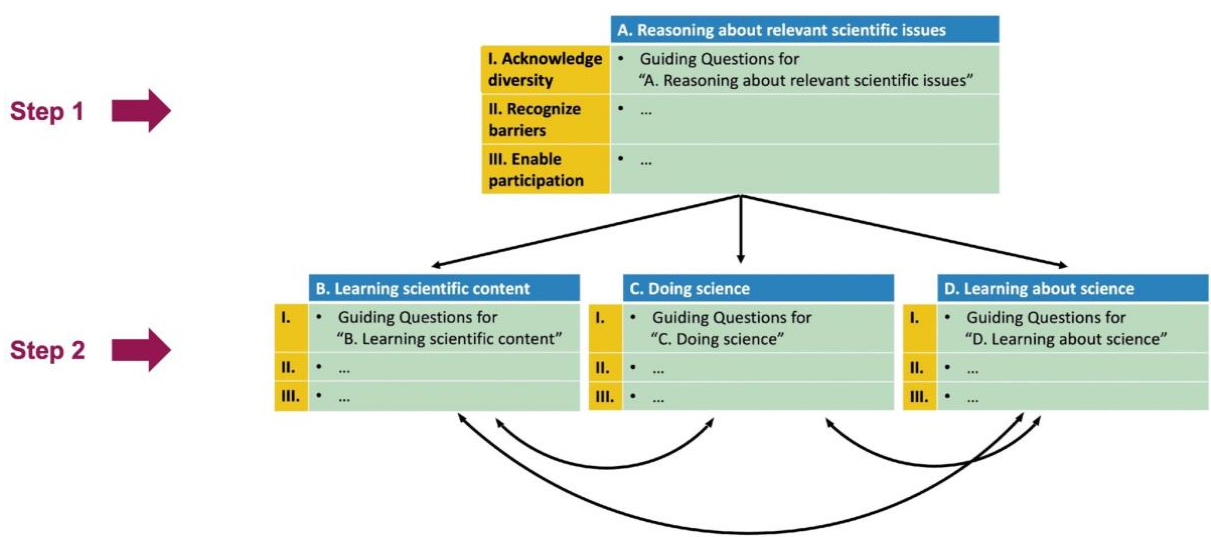

*Figure 2: Illustration of the application of the NinU-framework (adapted from Ferreira González et al., 2021).*

Step 1     Since scientific issues that are perceived as special or topical by learners can arouse their situational interest (Habig et al., 2018) and have a long-term meaningful effect on the science content taught (Gilbert et al, 2011), it is important to think about contextualizing the lesson first. Therefore, dealing with column *A. Reasoning about relevant scientific issues* should be seen as the basis for planning and reflecting on science lessons. Consequently, it is recommended that the guiding questions for A from top to bottom should be addressed first (see Fig. 2).

Step 2      Depending on the focus and goal of the lesson to be planned and reflected, another column (cf. Fig. 2: B, C, D) within the NinU-framework can be addressed for all guiding questions from top to bottom.

The remaining columns (cf. Fig. 2: B, C, D) can then be processed in the same way, if necessary. However, there is no requirement to cover all further columns (B, C, D) of the NinU-framework for each lesson. The columns B and C should be selected depending on the focus and goal of the lesson. This systematic approach to the use of the NinU-framework is intended to help teachers to keep all relevant aspects in mind and to support a conscious decision for certain learning goals of science education.

# IV. An Example "Making popcorn using a hair straightener"

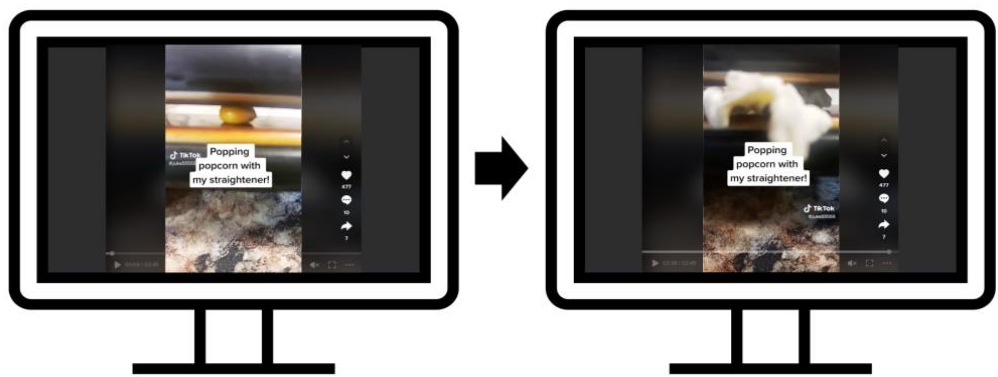

*Figure 3: Screenshots from a TikTok video by Julie Knight that is about making popcorn with a hair straightener. For the original video cf.: https://www.tiktok.com/@juke55555/video/7200914794344598830.*

The scientific issue presented in this self-study manual is the process of making popcorn using a hair straightener, as demonstrated in a TikTok video (cf. Fig. 3). From a subject-specific point of view this topic was chosen because it is relevant to various disciplines in science education, including biology, chemistry, and physics, and can be approached in an interdisciplinary manner. The context can be used in physics lessons to discuss pressure, temperature and heat transfer. In chemistry lessons the context can be used to thematize the molecular structure of starch and its denaturation. Addressing the structure of corn kernels can be interesting for biology lessons. From a pedagogical perspective, this context was chosen because it is relevant for different learners. The context can be experienced through multiple senses and is both interesting and motivating due to its relevance to everyday life.

An explanation of the phenomenon which can be used in a science lesson is provided in Figure 4.

An example solution for planning and reflecting a lesson with the context "making popcorn using a hair straightener" applying the NinU-framework can be found in the following chapter (V. Exercises & Examples). This chapter provides examples for all columns (A. Reasoning about relevant scientific issues, B. Learning scientific content, C. Doing science, D. Learning about science) and the included guiding questions of the framework.

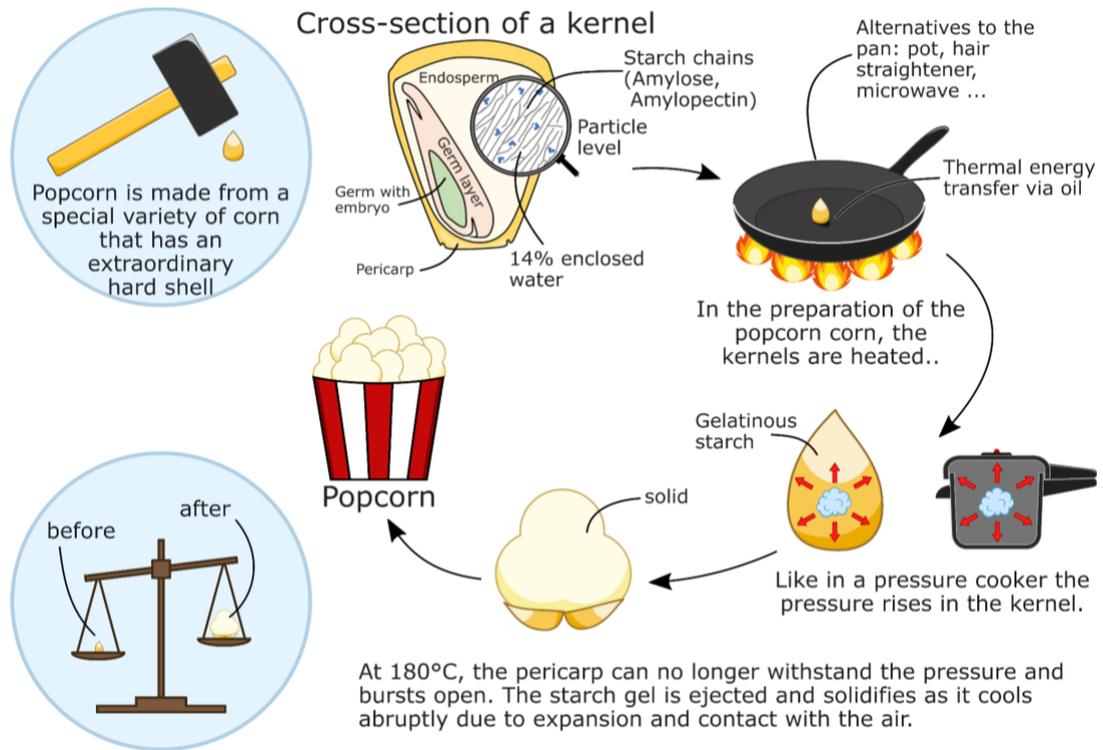

Figure 4: How popcorn is made *adapted from Ferreira González et al., 2021.*

Towards Scientific Literacy in Inclusive Science Education – A Self-Study Manual to Support Pre- and In-Service Teachers

# V. Exercises & Examples

## 1. Exercises

1. Read the complete example "Making popcorn with a hair straightener" for column **"A. Reasoning about scientific issues"** (cf. 1.1)

2. Use the NinU-Framework to plan & reflect the issue "Making popcorn with a hair straightener". Fill in the table for column **"B. Learning scientific content"** (cf. 1.2).

3. Use the provided self-study manual to compare your results with the example solution from the NinU-Network (cf. 1.3)
   Hint: this example solution from the NinU-Network can be used as a reference for Exercise 2.

S. Lenzer, L. Pannullo, A. Nehring & L. Stinken-Rösner                                                                                                              7

## 1.1 Example for "A. Reasoning about scientific issues"

|  | **Guiding Questions for "A. Reasoning about scientific issues"** | **Example for "A. Reasoning about scientific issues"** - **Popcorn popping" or What are the ways to make corn pop?** |
|---|---|---|
| **I. Acknowledge diversity** | 1. Which **scientific issues** are stimulating and relevant for all learners? | ● Popcorn machines in movie theaters can spark learners' interest in the context of popcorn making.<br>● Like a black box, it is not possible to see what is happening in the pot/boiler. It is not clear what products were put in to make it (e.g., oil) or are added afterwards (e.g., sugar).<br>● Videos (e.g., making popcorn with a hair straightener) can serve as a possible introductory phenomenon. |
|  | 2. Which dimensions of diversity play a role in **reasoning about scientific issues**? | ● It is possible that learners in year 7 are only interested in the context to a limited extent because, for example, it is not exciting enough.<br>● Gender may play a role, e.g., when the video with the hair straightener is used.<br>● Learners know popcorn machines from the cinema.<br>● Perceiving the phenomenon with all senses can influence interest.<br>● The consumption of popcorn differs. |
|  | 3. Which individual conceptions, skills, and beliefs of learners are related to **(reasoning about) the scientific issues**? | ● For example, learners might participate in class with the idea that popcorn is exclusively an industrial product.<br>● Perceptions of how popcorn is made (microwave, straightener, etc.) will vary. |
|  | 4. Which knowledge, skills, and experiences of learners can be seen as resources for **(reasoning about) the scientific issues**? | ● Learners know that popcorn can be made in popcorn machines, e.g., from county fairs or from the cinema.<br>● Learners have already tried to make popcorn in different ways (pot etc.). |
| **II. Recognize barriers** | 1. What are barriers and/or challenges for learners when **reasoning about scientific issues**? | <ul><li>The ambiguity of the term 'popping' can be challenging.</li><li>Stereotypes such as 'straighteners are a girls' thing' may be prevalent in the learning group.</li><li>Learners who are unfamiliar with popcorn lack the basic skills to engage with the context.</li><li>Learners might refuse to try homemade popcorn.</li><li>Fear of the popping sound/impaired sensory perception may limit interest.</li><li>Food intolerances may limit learners' encounters with popcorn.</li><li>Culturally, there may be an attitude that food preparation (e.g., popcorn) is "women's business."</li><li>The context of "What are some ways to make corn pop?" may present a distracting barrier (seductive details) for some learners through the specific question.</li></ul> |
| **III. Enable participation** | 1. How can **(reasoning about) the scientific issues** be made accessible to all learners? | <ul><li>Various approaches to the context are possible, e.g., tasting and smelling (basal-perceptive); exploring different characteristics (e.g. hardness, color, etc.) in an active way (concrete-active); looking at a picture/video of popcorn popping or making popcorn (vivid-pictorial); articulating experiences and making popcorn (symbolic-abstract).</li><li>Popcorn can be distributed to learners to eat, and the beginning of the TikTok video on "We make popcorn with a hair straightener" can be watched.</li></ul> |
| | 2. How can the existing resources be used to overcome the barriers or challenges when **reasoning about scientific issues**? | <ul><li>Learners express what they already know about the context of popcorn popping and share their experiences.</li><li>For learners who are unfamiliar with popcorn making methods, information is provided in the form of pictures, movies, etc.</li><li>Corn kernels are provided to make a comparison to popped corn.</li><li>Addressing the ambiguity of the term 'popping' as well as a laid-back approach helps to reduce barriers.</li><li>Gender stereotypes, such as only women cook, can be discussed, and critically reflected upon.</li></ul> |
| | 3. How can all learners be actively engaged when **reasoning about scientific issues**? | <ul><li>Provide appropriate stimuli, e.g., pictures of popcorn machines, feedback from teachers.</li><li>Tasks that involve all learners by asking them to position themselves, e.g., what their taste preference is or what machine they have already used to make popcorn.</li><li>Learners decide individually on a level of engagement (see III.A.1).</li></ul> |
| | 4. How can (all) learners be encouraged to co-construct and collaborate when **reasoning about scientific issues**? | ● Eating it can stimulate an exchange and lead to a co-constructive learning process through guiding questions.<br>● If learners are unfamiliar with popcorn, the example of others will make it clear to them that it can be eaten.<br>● An exchange, e.g., through think-pair-share procedures on popcorn production can be initiated.<br>● Any uncertainties about popcorn can be answered by fellow learners through this direct encounter. |
| | 5. How can all learners be individually supported when **reasoning about scientific issues**? | ● Learners who cannot visually observe the popcorn making process are given a spoken description of what is seen in the popcorn machine, TikTok video, or similar.<br>● If there is a fear of the popping sound, protective ear plugs can be offered, for example.<br>● Learners can be supported in communication, with talkers or by accompanying staff. |

## 1.2 Exercise for "B. Learning scientific content"

|  | **Guiding Questions for "B. Learning scientific content"** | **Exercise for "B. Learning scientific content - Popcorn"** |
|---|---|---|
| **I. Acknowledge diversity** | 1. Which **contents** are relevant for all learners? |  |
|  | 2. Which dimensions of diversity play a role in **learning the scientific content**? |  |
|  | 3. Which individual conceptions, skills, and beliefs of learners are related to **learning the scientific content**? |  |
|  | 4. Which knowledge, skills, and experiences of learners can be seen as resources for **learning the scientific** content? |  |
| **II. Recognize barriers** | 1. What are barriers and/or challenges for learners when **learning the scientific content**? | |
| **III. Enable participation** | 1. How can **(learning) the scientific content** be made accessible to all learners? | |
| | 2. How can the existing resources be used to overcome the barriers or challenges when **learning the scientific content**? | |
| | 3. How can all learners be actively engaged when **learning the scientific content**? | |
| | 4. How can (all) learners be encouraged to co-construct and collaborate when **learning the scientific content**? | |

Towards Scientific Literacy in Inclusive Science Education – A Self-Study Manual to Support Pre- and In-Service Teachers| | | |
|---|---|---|
| | 5. How can all learners be individually supported when **learning the scientific content**? | |

S. Lenzer, L. Pannullo, A. Nehring & L. Stinken-Rösner    13

## 1.3 Example solution for "B. Learning scientific content"

|  | **Guiding Questions for "B. Learning scientific content"** | **Example for "B. Learning scientific content - Popcorn"** |
|---|---|---|
| **I. Acknowledge diversity** | 1. Which **contents** are relevant for all learners? | ● Corn kernels as an exemplary starchy food or grain (e.g., IPM, 2020).<br>● Relationship between pressure, temperature and phase transitions of the water stored in the corn kernel<br>● Transfer of thermal energy => Changes of the corn kernel to the popped kernel (volume, mass, density)<br>● Molecular structure of starch, entrapment of water particles between starch chains; intermolecular interactions |
|  | 2. Which dimensions of diversity play a role in **learning the scientific content**? | ● The learners have different cognitive prerequisites, e.g., to grasp the process of popcorn production on an abstract level and the associated scientific content.<br>● The readiness to deal critically with content, e.g., what sugar content sweetened popcorn has, may be influenced by the circumstances at home. |
|  | 3. Which individual conceptions, skills, and beliefs of learners are related to **learning the scientific content**? | ● The concept of the structure of a corn kernel (micro level) and matter such as starch (submicroscopic level).<br>● The concepts of, for example, the difference between the states of matter and the transitions between them play an important role in learning the content.<br>● Understanding the concept of "pressure" and its relationship to heat, among other things, can also be crucial for grasping the content. |
|  | 4. Which knowledge, skills, and experiences of learners can be seen as resources for **learning the scientific** content? | ● Individual learners may already have prior knowledge, e.g., about …<br>   ○ …the structure of a corn kernel and knowledge of specific terminology.<br>   ○ …the structure of starch (e.g., from the potato) and/or the molecular structure of water.<br>   ○ …the heating (of the water in the kernel) leading to the bursting of the pericarp, the sound, and the reshaping of the kernel.<br>   ○ …the relationship between heat and expansion or pressure. |
| **II. Recognize barriers** | 1. What are barriers and/or challenges for learners when **learning the scientific content**? | ● It may not be possible for everyone to understand at an abstract level how the popped corn is created from the corn kernel.<br>● The causal chain (heat transfer => Expansion of water to steam => Ejection and change of starch), for example, may be too extensive or complex an event.<br>● The term popping is ambiguous. |
| **III. Enable participation** | 1. How can **(learning) the scientific content** be made accessible to all learners? | ● Different approaches, e.g., using a cut open corn kernel, offering a verbal description of the corn kernel.<br>● The previous knowledge is activated, e.g., by discussing the starch structure and transferring it to the potato. |
| | 2. How can the existing resources be used to overcome the barriers or challenges when **learning the scientific content**? | ● Show the transition from corn kernel to popcorn with a slow-motion video, delayed and enlarged.<br>● Visualize the substance level using illustrative materials (liquid water and water vapor).<br>● Illustrate the submicroscopic level using appropriate media, e.g., an animation showing the transition between liquid and gaseous states of matter and its relationship to heat.<br>● Word memory on which the everyday term (e.g., seed coat) and the technical term (e.g., pericarp) are noted.<br>● Use synonymous term to "popping" such as "bursting". |
| | 3. How can all learners be actively engaged when **learning the scientific content**? | ● Individual focus on content, e.g., on biological (structure of the corn kernel) or physical (change of aggregate state) contexts.<br>● Freedom of choice regarding the level of abstraction (e.g., role of water vapor on the level of phenomena or particles). |
| | 4. How can (all) learners be encouraged to co-construct and collaborate when **learning the scientific content**? | ● Discussion of content-related questions and/or exchange of individual ideas, explanations in cooperative ways (e.g., creation of a picture-based concept map).<br>● Present content (e.g., structure of a corn kernel, evaporation of water) through a self-selected presentation format. |
| | 5. How can all learners be individually supported when **learning the scientific content**? | ● Cognitive barriers and/or prolonged concentration on a content can be countered, e.g., by intensified action approaches (e.g., role play to illustrate the processes at particle level).<br>● Verbal exchange can be supported e.g., by computer/talkers or in the form of sign language interpreters. |

## 2. Exercises

1. Read the complete example "Making popcorn with a hair straightener" for column **"A. Reasoning about scientific issues"** (cf. 2.1)

2. Use the NinU-Framework to plan & reflect the issue "Making popcorn with a hair straightener". Fill in the table for column **"C. Doing science"** (cf. 2.2).

3. Use the provided self-study manual to compare your results with the example solution from the NinU-Network (cf. 2.3)
   Hint: this example solution from the NinU-Network can be used as a reference for Exercise 2.

## 2.1 Example for "A. Reasoning about scientific issues"

|  | **Guiding Questions for "A. Reasoning about scientific issues"** | **Example for "A. Reasoning about scientific issues"** - **Popcorn popping" or What are the ways to make corn pop?** |
|---|---|---|
| **I. Acknowledge diversity** | 1. Which **scientific issues** are stimulating and relevant for all learners? | ● Popcorn machines in movie theaters can spark learners' interest in the context of popcorn making.<br>● Like a black box, it is not possible to see what is happening in the pot/boiler. It is not clear what products were put in to make it (e.g., oil) or are added afterwards (e.g., sugar).<br>● Videos (e.g., making popcorn with a hair straightener) can serve as a possible introductory phenomenon. |
|  | 2. Which dimensions of diversity play a role in **reasoning about scientific issues**? | ● It is possible that learners in year 7 are only interested in the context to a limited extent because, for example, it is not exciting enough.<br>● Gender may play a role, e.g., when the video with the hair straightener is used.<br>● Learners know popcorn machines from the cinema.<br>● Perceiving the phenomenon with all senses can influence interest.<br>● The consumption of popcorn differs. |
|  | 3. Which individual conceptions, skills, and beliefs of learners are related to **(reasoning about) the scientific issues**? | ● For example, learners might participate in class with the idea that popcorn is exclusively an industrial product.<br>● Perceptions of how popcorn is made (microwave, straightener, etc.) will vary. |
|  | 4. Which knowledge, skills, and experiences of learners can be seen as resources for **(reasoning about) the scientific issues**? | ● Learners know that popcorn can be made in popcorn machines, e.g., from county fairs or from the cinema.<br>● Learners have already tried to make popcorn in different ways (pot etc.). |
| **II. Recognize barriers** | 1. What are barriers and/or challenges for learners when **reasoning about scientific issues**? | ● The ambiguity of the term 'popping' can be challenging.<br>● Stereotypes such as 'straighteners are a girls' thing' may be prevalent in the learning group.<br>● Learners who are unfamiliar with popcorn lack the basic skills to engage with the context.<br>● Learners might refuse to try homemade popcorn.<br>● Fear of the popping sound/impaired sensory perception may limit interest.<br>● Food intolerances may limit learners' encounters with popcorn.<br>● Culturally, there may be an attitude that food preparation (e.g., popcorn) is "women's business."<br>● The context of "What are some ways to make corn pop?" may present a distracting barrier (seductive details) for some learners through the specific question. |
| **III. Enable participation** | 1. How can **(reasoning about) the scientific issues** be made accessible to all learners? | ● Various approaches to the context are possible, e.g., tasting and smelling (basal-perceptive); exploring different characteristics (e.g., hardness, color, etc.) in an active way (concrete-active); looking at a picture/video of popcorn popping or making popcorn (vivid-pictorial); articulating experiences and making popcorn (symbolic-abstract).<br>● Popcorn can be distributed to learners to eat, and the beginning of the TikTok video on "We make popcorn with a hair straightener" can be watched. |
| | 2. How can the existing resources be used to overcome the barriers or challenges when **reasoning about scientific issues**? | ● Learners express what they already know about the context of popcorn popping and share their experiences.<br>● For learners who are unfamiliar with popcorn making methods, information is provided in the form of pictures, movies, etc.<br>● Corn kernels are provided to make a comparison to popped corn.<br>● Addressing the ambiguity of the term 'popping' as well as a laid-back approach helps to reduce barriers.<br>● Gender stereotypes, such as only women cook, can be discussed, and critically reflected upon. |
| | 3. How can all learners be actively engaged when **reasoning about scientific issues**? | ● Provide appropriate stimuli, e.g., pictures of popcorn machines, feedback from teachers.<br>● Tasks that involve all learners by asking them to position themselves, e.g., what their taste preference is or what machine they have already used to make popcorn.<br>● Learners decide individually on a level of engagement (see III.A.1). |
| | 4. How can (all) learners be encouraged to co-construct and collaborate when **reasoning about scientific issues**? | ● Eating it can stimulate an exchange and lead to a co-constructive learning process through guiding questions.<br>● If learners are unfamiliar with popcorn, the example of others will make it clear to them that it can be eaten.<br>● An exchange, e.g., through think-pair-share procedures on popcorn production can be initiated.<br>● Any uncertainties about popcorn can be answered by fellow learners through this direct encounter. |
| | 5. How can all learners be individually supported when **reasoning about scientific issues**? | ● Learners who cannot visually observe the popcorn making process are given a spoken description of what is seen in the popcorn machine, TikTok video, or similar.<br>● If there is a fear of the popping sound, protective ear plugs can be offered, for example.<br>● Learners can be supported in communication, with talkers or by accompanying staff. |

## 2.2 Exercise for "C. Doing science"

|  | **Guiding questions for "C. Doing science"** | **Exercise for "C. Doing science"** |
|---|---|---|
| **I. Acknowledge diversity** | 1. Which **processes and procedures of doing science** are relevant for all learners? |  |
|  | 2. Which dimensions of diversity play a role for **doing science**? |  |
|  | 3. Which individual conceptions, skills, and beliefs of learners are related to **doing science**? |  |
|  | 4. Which knowledge, skills, and experiences of learners can be seen as resources for **doing science**? |  |
| **II. Recognize barriers** | 1. What are barriers and/or challenges for learners when **doing science**? | |
| **III. Enable participation** | 1. How can **doing science** be made accessible to all learners? | |
| | 2. How can the existing resources be used to overcome the barriers or challenges when **doing science**? | |
| | 3. How can all learners be actively engaged when **doing science**? | |
| | 4. How can (all) learners be encouraged to co-construct and collaborate when **doing science**? | |

|  | 5. How can all learners be individually supported when **doing science**? |  |
|---|---|---|

## 2.3 Example solution for "C. Doing science"

|  | **Guiding questions for "C. Doing science"** | **Example solution for "C. Doing science"** |
|---|---|---|
| **I. Acknowledge diversity** | 1. Which **processes and procedures of doing science** are relevant for all learners? | ● Working with models using a model at the particle level, e.g., how water is intercalated between the starch chains.<br>● Microscopy by looking at the cross-section of the kernel and popcorn under the microscope.<br>● Observing changes in the corn kernel compared to the popped kernel (volume, mass, density). |
|  | 2. Which dimensions of diversity play a role for **doing science**? | ● What materials are the learners allowed and willing to experiment with (smoothing irons, waffle irons, etc.)?<br>● Learners bring different experiences, e.g., in handling food or electrical appliances.<br>● Learners bring different relevant motor skills for handling corn kernel.<br>● Learners have different sensory perceptions (e.g., hearing, seeing) to perform experiments or perceive results (popping sound audible). |
|  | 3. Which individual conceptions, skills, and beliefs of learners are related to **doing science**? | ● Learners already have experience in using the electrical equipment provided.<br>● Learners might know different ways of preparation.<br>● Different pre-concepts ("making popcorn works with any corn", "the corn kernel consists of a husk and is filled with starch") are relevant for engaging in scientific knowledge discovery. |
|  | 4. Which knowledge, skills, and experiences of learners can be seen as resources for **doing science**? | ● Learners know the safety rules when experimenting with electrical and heat-generating devices.<br>● Learners have made popcorn before and know what to look for (e.g., that sufficient heat is transferred, the appropriate type of corn is selected).<br>● Learners have a model idea of the structure of matter from the smallest particles. |
| **II. Recognize barriers** | 1. What are barriers and/or challenges for learners when **doing science**? | ● Certain heat-making equipment may be unfamiliar and thus it can be challenging to use it safely.<br>● Learners who are unfamiliar with the equipment presented and its usage for making popcorn may feel overwhelmed by the situation. They may also feel inhibited by/in front of other co-learners.<br>● Cognitive skills can become a barrier when the functionality of certain popcorn making equipment and its operation cannot be grasped.<br>● Cognitive skills can become a barrier when there is an overload of experimental skills, such as understanding and sustaining a variable-controlled procedure in which only the dependent variables (e.g., temperature, time) are changed. |
| **III. Enable participation** | 1. How can **doing science** be made accessible to all learners? | ● Different levels, appropriation levels and approaches could be e.g.: listening to the popping sound, experiencing the difference between the tastes by smelling/tasting (basal-perceptive); making popcorn as an introductory experiment, e.g. in the microwave/boiling pot, or practically pursue different experimental approaches (concrete-acting); watch a comic or video on the topic of "the crazy professor popping popcorn" or look at a model representation of the processes at the particle level (vivid-pictorial); represent the relationship between temperature and the number of popping corn kernels in the form of a diagram (symbolic-abstract). |
| | 2. How can the existing resources be used to overcome the barriers or challenges when **doing science**? | ● Learners can look for equipment to make popcorn in the school (e.g., school kitchen).<br>● Various experimental materials can be prepared and made available, e.g., waffle and smoothing irons, stove top, cooking pots and other heat resistant vessels, microwave, corn varieties.<br>● Various observation opportunities are provided, e.g., a SloMo camera. |
| | 3. How can all learners be actively engaged when **doing science**? | ● All learners formulate assumptions about a selected question, e.g., which device makes corn pop the "fastest."<br>● There are enough electrical devices available for popcorn production so that several small groups can work at the same time. |
| | 4. How can (all) learners be encouraged to co-construct and collaborate when **doing science**? | ● Create positive interdependencies in small groups, e.g., in the case of the question on the intensity of noise development, each member of the small group can carry out an experiment/series of measurements in which a possible characteristic (e.g., ovenproof glass jar, stainless steel pot) is investigated and only the joint comparison of all results leads to a solution. |
| | 5. How can all learners be individually supported when **doing science**? | ● If the function of certain devices and their operation cannot be grasped, others (e.g., fellow learners) can assist with the application or assistive aids (e.g., Powerlink) can be used.<br>● In the case of affective barriers, such as fear of electricity or heat, watching can be made possible and a gradual introduction to the experimentation process with electricity or heat can take place. |

Towards Scientific Literacy in Inclusive Science Education – A Self-Study Manual to Support Pre- and In-Service Teachers## 3. Exercises

1. Read the complete example "Making popcorn with a hair straightener" for column **"A. Reasoning about scientific issues"** (cf. 3.1).

2. Use the NinU-Framework to plan & reflect the issue "Making popcorn with a hair straightener". Fill in the table for column **"D. Learning about science"** (cf. 3.2).

3. Use the provided self-study manual to compare your results with the example solution from the NinU-Network.
   Hint: this example solution from the NinU-Network can be used as a reference for Exercise 2 (3.3).

S. Lenzer, L. Pannullo, A. Nehring & L. Stinken-Rösner		27

## 3.1 Example for "A. Reasoning about scientific issues"

|  | Guiding Questions for "A. Reasoning about scientific issues" | Example for "A. Reasoning about scientific issues" - **Popcorn popping" or What are the ways to make corn pop?** |
|---|---|---|
| **I. Acknowledge diversity** | 1. Which **scientific issues** are stimulating and relevant for all learners? | • Popcorn machines in movie theaters can spark learners' interest in the context of popcorn making.<br>• Like a black box, it is not possible to see what is happening in the pot/boiler. It is not clear what products were put in to make it (e.g., oil) or are added afterwards (e.g., sugar).<br>• Videos (e.g., making popcorn with a hair straightener) can serve as a possible introductory phenomenon. |
|  | 2. Which dimensions of diversity play a role in **reasoning about scientific issues**? | • It is possible that learners in year 7 are only interested in the context to a limited extent because, for example, it is not exciting enough.<br>• Gender may play a role, e.g., when the video with the hair straightener is used.<br>• Learners know popcorn machines from the cinema.<br>• Perceiving the phenomenon with all senses can influence interest.<br>• The consumption of popcorn differs. |
|  | 3. Which individual conceptions, skills, and beliefs of learners are related to **(reasoning about) the scientific issues**? | • For example, learners might participate in class with the idea that popcorn is exclusively an industrial product.<br>• Perceptions of how popcorn is made (microwave, straightener, etc.) will vary. |
|  | 4. Which knowledge, skills, and experiences of learners can be seen as resources for **(reasoning about) the scientific issues**? | • Learners know that popcorn can be made in popcorn machines, e.g., from county fairs or from the cinema.<br>• Learners have already tried to make popcorn in different ways (pot etc.). |
| **II. Recognize barriers** | 1. What are barriers and/or challenges for learners when **reasoning about scientific issues**? | <ul><li>The ambiguity of the term 'popping' can be challenging.</li><li>Stereotypes such as 'straighteners are a girls' thing' may be prevalent in the learning group.</li><li>Learners who are unfamiliar with popcorn lack the basic skills to engage with the context.</li><li>Learners might refuse to try homemade popcorn.</li><li>Fear of the popping sound/impaired sensory perception may limit interest.</li><li>Food intolerances may limit learners' encounters with popcorn.</li><li>Culturally, there may be an attitude that food preparation (e.g., popcorn) is "women's business."</li><li>The context of "What are some ways to make corn pop?" may present a distracting barrier (seductive details) for some learners through the specific question.</li></ul> |
| **III. Enable participation** | 1. How can **(reasoning about) the scientific issues** be made accessible to all learners? | <ul><li>Various approaches to the context are possible, e.g., tasting and smelling (basal-perceptive); exploring different characteristics (e.g., hardness, color, etc.) in an active way (concrete-active); looking at a picture/video of popcorn popping or making popcorn (vivid-pictorial); articulating experiences and making popcorn (symbolic-abstract).</li><li>Popcorn can be distributed to learners to eat, and the beginning of the TikTok video on "We make popcorn with a hair straightener" can be watched.</li></ul> |
| | 2. How can the existing resources be used to overcome the barriers or challenges when **reasoning about scientific issues**? | <ul><li>Learners express what they already know about the context of popcorn popping and share their experiences.</li><li>For learners who are unfamiliar with popcorn making methods, information is provided in the form of pictures, movies, etc.</li><li>Corn kernels are provided to make a comparison to popped corn.</li><li>Addressing the ambiguity of the term 'popping' as well as a laid-back approach helps to reduce barriers.</li><li>Gender stereotypes, such as only women cook, can be discussed, and critically reflected upon.</li></ul> |
| | 3. How can all learners be actively engaged when **reasoning about scientific issues**? | <ul><li>Provide appropriate stimuli, e.g., pictures of popcorn machines, feedback from teachers.</li><li>Tasks that involve all learners by asking them to position themselves, e.g., what their taste preference is or what machine they have already used to make popcorn.</li><li>Learners decide individually on a level of engagement (see III.A.1).</li></ul> |
| | 4. How can (all) learners be encouraged to co-construct and collaborate when **reasoning about scientific issues**? | ● Eating it can stimulate an exchange and lead to a co-constructive learning process through guiding questions.<br>● If learners are unfamiliar with popcorn, the example of others will make it clear to them that it can be eaten.<br>● An exchange, e.g., through think-pair-share procedures on popcorn production can be initiated.<br>● Any uncertainties about popcorn can be answered by fellow learners through this direct encounter. |
| | 5. How can all learners be individually supported when **reasoning about scientific issues**? | ● Learners who cannot visually observe the popcorn making process are given a spoken description of what is seen in the popcorn machine, TikTok video, or similar.<br>● If there is a fear of the popping sound, protective ear plugs can be offered, for example.<br>● Learners can be supported in communication, with talkers or by accompanying staff. |

## 3.2 Exercise for "D. Learning about science"

|  | **Guiding questions for "D. Learning about science"** | **Exercise for "D. Learning about science"** |
|---|---|---|
| **I. Acknowledge diversity** | 1. Which **aspects of learning about science** are relevant for all learners? | |
|  | 2. Which dimensions of diversity play a role for **learning about science**? | |
|  | 3. Which individual conceptions, skills, and beliefs of learners are related to **learning about science**? | |
|  | 4. Which knowledge, skills, and experiences of learners can be seen as resources for **learning about science**? | |
| **II. Recognize barriers** | 1. What are barriers and/or challenges for learners when **learning about science**? | |
| **III. Enable participation** | 1. How can **learning about science** be made accessible to all learners? | |
| | 2. How can the existing resources be used to overcome the barriers or challenges when **learning about science**? | |
| | 3. How can all learners be actively engaged when **learning about science**? | |
| | 4. How can (all) learners be encouraged to co-construct and collaborate when **learning about science**? | |
| | 5. How can all learners be individually supported when **learning about science**? | |

## 3.3 Example solution for "D. Learning about science"

|  | Guiding questions for "D. Learning about science" | Example solution for "D. Learning about science" |
|---|---|---|
| **I. Acknowledge diversity** | 1. Which **aspects of learning about science** are relevant for all learners? | ● Learners can understand that research builds on previous knowledge. Virot and Ponomarenko (2015) show that earlier studies have focused on the conditions needed for successful popping, and they address the rupture of the pericarp, where questions remain.<br>● Researchers publish their findings in a variety of ways (e.g., papers, YouTube).<br>● The temperature optimum experiment could be repeated endlessly and lead to wasted corn.<br>● The social benefit of popcorn research is particularly for the industrial preparation of popcorn. |
|  | 2. Which dimensions of diversity play a role for **learning about science**? | ● The age of the learners influences which media they are legally allowed or able to use.<br>● In the example chosen, two supposedly male scientists present their results.<br>● The understanding of values regarding the use of resources can vary ("Am I allowed to use food like corn for my research?"). |
|  | 3. Which individual conceptions, skills, and beliefs of learners are related to **learning about science**? | ● Learners may have the idea that- scientific findings are to be equated with absolute facts and are not open to discussion vs. can always be discussed (e.g., the popping sound is caused by the pericarp breaking).<br>● Results can always be generalized ("cooking in the pot only works with oil").<br>● Experiments must always prove theories (popping rate: number of pops per unit time is normally distributed). |
|  | 4. Which knowledge, skills, and experiences of learners can be seen as resources for **learning about science**? | ● Learners are aware that scientific findings can trigger social progress (e.g., through knowledge of the optimum temperature in industrial popcorn production).<br>● Learners who have experience in experimentation - can understand how lengthy a research process can be (e.g., experiment on the optimum temperature).<br>● Learners who have experience in experimentation know that scientists can deal constructively with unexpected results and that it can be a reason to continue research ("The recordings have shown that the popping noise is not associated with the rupture of the pericarp, so we are investigating how and when the noise occurs"). |
| **II. Recognize barriers** | 1. What are barriers and/or challenges for learners when **learning about science**? | ● The age of the learners may limit the free search on the Internet about the scientists and scientific studies on basic rights.<br>● Girls may be discouraged by the example of the two male scientists.<br>● Learners who do not consider the value of resources may underestimate this social responsibility. |
| **III. Enable participation** | 1. How can **learning about science** be made accessible to all learners? | ● Role play of a conference, presentation of the results from column C, can support access to the meta level.<br>● View pictures or videos of research processes and results presentations of scientists5 (vivid pictorial).<br>● Learners could conduct interviews with scientists in their environment and thereby learn something about the everyday work and their results, e.g., in relation to popcorn (symbolic-abstract). |
| | 2. How can the existing resources be used to overcome the barriers or challenges when **learning about science**? | ● Learners who have experience can report on it.<br>● Learners could inform themselves on the Internet (texts, videos, etc.) about the activities of scientists, especially in the food industry, on the topic of popcorn and, if necessary, be supported by personnel.<br>● The process of gaining scientific knowledge could be explicitly illustrated using the example of Virot and Ponomarenko's (2015) research on the optimum temperature and discussed with the learners. |
| | 3. How can all learners be actively engaged when **learning about science**? | ● Learners can take on different roles in a laboratory setting (e.g., a series on taste testing popcorn, see III.D.1), such as laboratory manager, technical assistant, consumer.<br>● Together with the learners, they consider and discuss what constitutes responsible use (waste, research with food, specifically corn kernels) in the research process. |
| | 4. How can (all) learners be encouraged to co-construct and collaborate when **learning about science**? | ● The learners prepare the experiments carried out in the phase of gaining knowledge and the results obtained scientifically, e.g., by working in small groups to produce a video, poster, etc.<br>● Everyone is given the task of actively contributing to at least one other group in the form of feedback. |
| | 5. How can all learners be individually supported when **learning about science**? | ● Role descriptions for all/particular learners can help develop reasoning strategies.<br>● Learners who want a more in-depth look could explore further aspects such as cost factor, time factor, etc. of research projects. |

# VI. References & Additional Literature

## VII. Acknowledgements

We would like to express our gratitude to Prof. Dr. Simone Abels for supporting this workshop. Additionally, we extend our thanks to all members of the NinU-network. Without their discussions and preparatory contributions, this self-study manual would not have been possible.

## VIII. License

A self-study manual by Stefanie Lenzer, Laura Pannullo, Andreas Nehring & Lisa Stinken-Rösner

supported by the NinU-Network.

This self-study manual is licensed under:

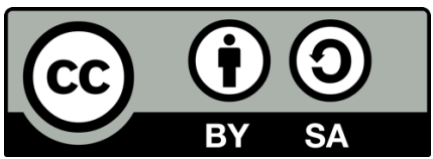

© Stefanie Lenzer, Laura Pannullo, Andreas Nehring & Lisa Stinken-Rösner, 2024. This manual is released under the Creative Commons Attribution-ShareAlike, version 4.0 International (CC BY-SA 4.0 en).
URL: https://creativecommons.org/licenses/by-sa/4.0/

## VIII. Contact

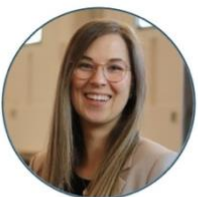
Dr. Stefanie Lenzer
Leibniz Institute for Science and Mathematics Education

lenzer@leibniz-ipn.de

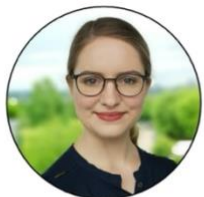
Laura Pannullo
University Bielefeld

laura.pannullo
@physik.uni-bielefeld.de

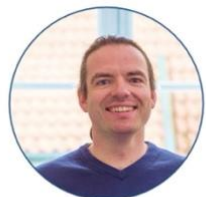
Prof. Dr. Andreas Nehring
Leibniz University Hannover

nehring@idn.uni-hannover.de

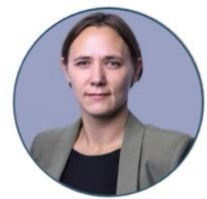
Prof. Dr. Lisa Stinken-Rösner
University Bielefeld

lisa.stinken-roesner
@physik.uni-bielefeld.de

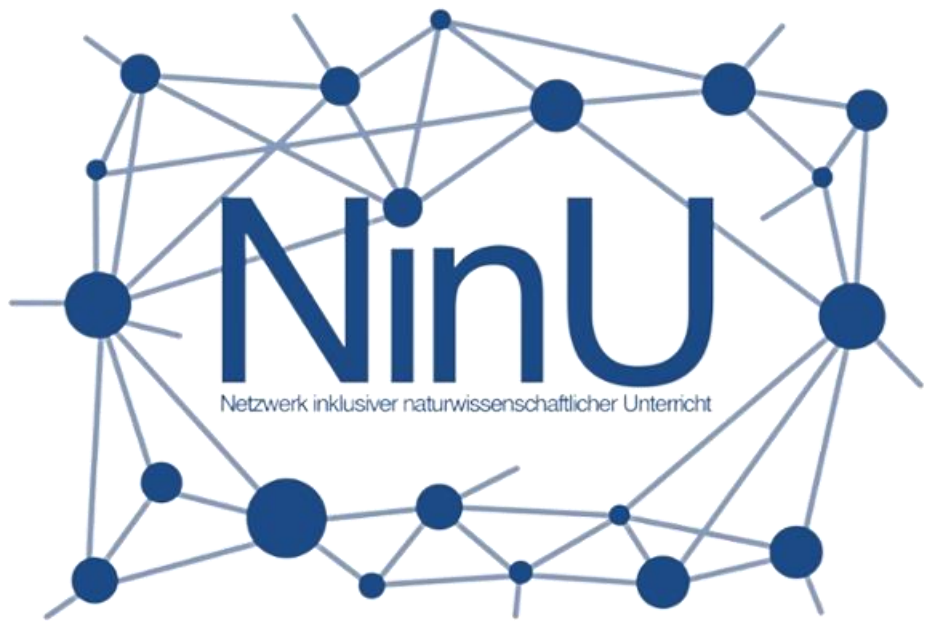